%% file: main.tex
\documentclass[manuscript]{acmart}
\renewcommand\footnotetextcopyrightpermission[1]{} % removes footnote with conference information in first column
\AtBeginDocument{%
  \providecommand\BibTeX{{%
    \normalfont B\kern-0.5em{\scshape i\kern-0.25em b}\kern-0.8em\TeX}}}

\setcopyright{acmcopyright}
\copyrightyear{2021}
\acmYear{2021}
\acmDOI{}

\acmConference[EASE'21]{EASE'21: International Conference on Evaluation and Assessment in Software Engineering}{June 21--23, 2021}{Trondheim, Norway}
\acmBooktitle{EASE'21: International Conference on Evaluation and Assessment in Software Engineering,
  June 21--23, 2021, Trondheim, Norway}
\acmPrice{15.00}
\acmISBN{978-1-4503-XXXX-X/18/06}

\begin{document}

\title{Storytelling in human--centric software engineering research}

\author{Austen Rainer}
\email{a.rainer@qub.ac.uk}
\orcid{0000-0001-8868-263X}
\affiliation{%
  \institution{Queen's University Belfast}
  \streetaddress{School of Electronics, Electrical Engineering and Computer Science}
  \city{Belfast}
  \state{County Antrim}
  \country{Northern Ireland}
  \postcode{BT9 6SB}
}

% \affiliation{%

% \affiliation{%

% \affiliation{%

% \affiliation{%

% \affiliation{%

% \affiliation{%

% \affiliation{%

\renewcommand{\shortauthors}{Rainer}

\input{sections/0_abstract}
% \begin{abstract}
%   A clear and well-documented \LaTeX\ document is presented as an
%   article formatted for publication by ACM in a conference proceedings
%   or journal publication. Based on the ``acmart'' document class, this
%   article presents and explains many of the common variations, as well
%   as many of the formatting elements an author may use in the
%   preparation of the documentation of their work.
% \end{abstract}




\maketitle

\input{sections/1_introduction}
\input{sections/2_foundations}

\input{sections/3_contribution_of_storytelling}
\input{sections/4_discussion}
\begin{acks}
We thank Dr Maria Angela Ferrario for her comments on an early version of this paper, and the anonymous reviewers for their constructive feedback.
\end{acks}

\bibliographystyle{ACM-Reference-Format}
\bibliography{sample-base}

% \appendix









\end{document}

%% file: sections/0_abstract.tex
\begin{abstract}
BACKGROUND: Software engineering is a human activity. People naturally make sense of their activities and experience through storytelling. But storytelling does not appear to have been properly studied by software engineering research.\\
AIM:  We explore the question: what contribution can storytelling make to human--centric software engineering research?\\ 
METHOD: We define concepts, identify types of story and their purposes, outcomes and effects, briefly review prior literature, identify several contributions and propose next steps.\\
RESULTS: Storytelling can, amongst other contributions, contribute to data collection, data analyses, ways of knowing, research outputs, interventions in practice, and advocacy, and can integrate with evidence and arguments. Like all methods, storytelling brings risks. These risks can be managed.\\ 
CONCLUSION: Storytelling provides a potential counter--balance to abstraction, and an approach to retain and honour human meaning in software engineering.\\
\end{abstract}

%% file: sections/1_introduction.tex
\begin{flushright}
% \begin{displayquote}
$ \sim $ If story is central to human meaning why, in the research world, is there not more storytelling? \cite{lewis2011storytelling}
% \end{displayquote}
\end{flushright}

\section{Introduction}
\label{section:introduction}

Software engineers generate and transform abstractions but they are not themselves abstractions. Software engineers have the same human characteristics as non--software engineers. They have goals \cite{van2001goal} and intentions \cite{ghapanchi2011antecedents, lenberg2015behavioral}; they wrestle with motivation \cite{beecham2008motivation}, happiness \cite{graziotin2017consequences}, stress \cite{ostberg2020methodology}, politics \cite{bergman2002large}, ethics \cite{singer2002ethical} and human values \cite{winter2018measuring}. They work individually, and in teams, projects, organisations, the wider software industry, and society \cite{curtis1988field, defranco2017review}. The software engineer's work with abstractions therefore takes place within the common sphere of human activity and experience. In other words, software engineering (SE) is a human--centric activity \cite{grundy2020towards, lenberg2015behavioral}.

Haidt writes that the human mind is a story processor not a logic processor (\cite{haidt2012righteous}, p. 328) and that, ``\dots [a]mong the most important stories we know are stories about ourselves\dots'' (\cite{haidt2012righteous}, p. 328). Storytelling would therefore be a natural way for software engineers to make sense of their own and other's behaviour, and for researchers to better understand this human--centric activity.

% Human--centric software engineering activity occurs in layers: software engineers necessarily communicate with each other individually, and in teams, projects, organisations, the wider software industry, and society \cite{curtis1988field, defranco2017review}.

% People tell stories to make sense of their worlds. Those worlds include the behaviour of others. Such stories will likely be found wherever people communicate with each other, professionally or otherwise. 

% Software engineers are people who necessarily communicate with each other individually, and in teams, projects, organisations, the wider software industry, and society \cite{curtis1988field, defranco2017review}. Communication occurs throughout the entire software engineering lifecycle, e.g., from software requirements elicitation (and the user story) to information--sharing between software engineers, to reporting progress and status to senior managers, to the sharing of experience with peers via social media, to providing information to researchers in empirical studies. 

Given the above, this paper explores the following
% we seek to better understand the contribution that storytelling can make to human--centric software engineering (SE). The paper explores the following,
simply--stated question: what contribution can storytelling make to human--centric software engineering research? The paper identifies several opportunities for storytelling to contribute to software engineering research, recognises there are risks (that can be managed) in using storytelling, and proposes next steps.

The remainder of this paper is organised as follows: in Section \ref{section:foundations} we provide foundations, including concepts, challenges in SE research and a brief review of prior research; in Section \ref{section:contribution-of-storytelling} we discuss several ways in which storytelling can contribute to SE research; and in Section \ref{section:discussion} we briefly consider next steps, and then conclude.

% Junk

% \subsection*{Notes}
% What is the place of story and storytelling in software engineering?
% \begin{enumerate}
%     \item Story and storytelling has been used effectively in other disciplines.
%     \item There is extensive use of a particular kind of `story' in software engineering practice (i.e., the user story).
%     \item There is some appreciation of story in software engineering research, e.g., war stories.
%     \item One previous paper -- Sims et al. -- approach story and storytelling from the perspective we do.
%     \item One unpublished manuscript: Lenberg et al.
%     \item There are challenges in SE: can story and storytelling help?
%     \item There are differences between storytelling and narrative
% \end{enumerate}

% \subsection*{Introduction}

%% file: sections/2_foundations.tex
\section{Foundations}
\label{section:foundations}

% \subsection*{Notes}
% \begin{enumerate}
%     \item Define what we mean by story, distinguishing it from narrative
%     \item Elements of story
%     \begin{enumerate}
%         \item A performative act
%     \end{enumerate}
%     \item Introduce and discuss model/s of story, e.g.,
%     \begin{itemize}
%         \item behavioural medicine Shaffer et al., 
%         \item law
%     \end{itemize}
%     \item Immersion connects to good storytelling
%     \item Recognise dangers with story and storytelling
%     \begin{itemize}
%         \item Parnas and confabulation
%     \end{itemize}
% \end{enumerate}

\subsection{Concepts}

% \subsubsection{Definitions}

There are many definitions for the terms \textit{story}, \textit{storytelling} and \textit{narrative}, and disciplines use other terms too. For example, Shaffer et al. \cite{shaffer2018usefulness} observe that journalism uses the term \textit{exemplar}, marketing uses \textit{testimonial}, psychology uses \textit{case study} or \textit{case history}, sociologists use \textit{personal stories} and medicine uses \textit{narrative}. Polkinghorne \cite{polkinghorne1995narrative} explores contrasting definitions of the term \textit{narrative}, distinguishing between the following: narrative as a prosaic discourse differentiated from poetic discourse, narrative as the qualitative inquiry into naturally produced linguistic expressions in their context, narrative as the collected body of data for analysis, narrative as the research output from qualitative inquiry, and narrative as a special type of discourse production, i.e., the story. We use Polkinghorne's \cite{polkinghorne1995narrative} definition/s of storytelling, summarised in Table \ref{table:definitions_of_story}.

It can be convenient to distinguish between story and storytelling, where story is the (static) output from a storytelling process. An exemplar of story is the novel. Although the story may in some sense be static, the story remains a dynamic interaction between the output and the reader. So although in some sense a printed story is fixed it is available as input to a re--telling process: the story\textit{telling} continues in the reading of the story. Digressing briefly, this suggests an intriguing metaphor for software: software is static until it is read -- `retold' -- by a processor.

\begin{table*}
  \caption{Example descriptions of storied narrative (from Polkinghorne \cite{polkinghorne1995narrative})}
  \label{table:definitions_of_story}
  \begin{tabular}{l p{13cm}}
    \toprule
    \# & Description\\
    \midrule
    1 & ``A story is a special type of discourse production. In a story, events and actions are drawn together into an organized whole by means of a plot. A plot is a type of conceptual scheme by which a contextual meaning of individual events can be displayed.'' (\cite{polkinghorne1995narrative}, p. 7)\\
    2 & ``\dots a specific kind of prose text (the story) and\dots the particular kind of configuration that generates a story (emplotment).'' (p. 5)  \\
    3 & ``A storied narrative is the linguistic form that preserves the complexity of human action with its interrelationship of temporal sequence, human motivation, chance happenings, and changing interpersonal and environmental contexts. In this context, \textit{story} refers not only to fictional accounts but also to narratives describing `ideal' life events such as biographies, autobiographies, histories, case studies, and reports of remembered episodes that have occurred.'' (\cite{polkinghorne1995narrative}, p. 7; emphasis in original), and\\
    4 & ``The subject--matter of stories is human action. Stories are concerned with human attempts to progress to a solution, clarification, or unraveling of an incomplete situation. '' (\cite{polkinghorne1995narrative}, p. 7)\\
  \bottomrule
\end{tabular}
\end{table*}

% \subsubsection{Elements}

Just as researchers and story--analysts do not agree on definitions for narrative and for story, so there is disagreement on the necessary elements of a story narrative. Indicative elements of storytelling include: a plot, a narrative structure, one or more protagonists, one or more antagonists, conflict, inciting incident/s, locations in time/s and space/s, and `space' for the reader to `inhabit' the story. (See, for example, \cite{coyne2015story,storr2020science} for further information.)

Shaffer and Zikmund--Fisher \cite{shaffer2013all} identify several purposes and outcomes for storied narrative, and Shaffer et al. \cite{shaffer2018usefulness} identify nine effects of narratives. These purposes, outcomes and effects are summmarised in Table \ref{table:purpose_outcome_effects_of_story}. As the table indicates, storytelling has many outcomes relevant to software engineering practice and research beyond entertainment.

\begin{table}
  \caption{Purpose, outcome \cite{shaffer2013all}, \& effect \cite{shaffer2018usefulness} of storytelling}
  \label{table:purpose_outcome_effects_of_story}
  \begin{tabular}{l p{2.6cm} p{4.2cm}}
    \toprule
    \# & Purpose & Outcome\\
    \midrule
    a & To inform & Improved knowledge\\
      & & Improved affective forecasting \\
    b & To engage & Greater engagement\\
      & & Greater transportation \\
      & & Greater time spent with materials \\
    c & To model behaviour & Increased participation\\
      & & Increased shared decision making \\
      & & Altered behavioural intentions \\
      & & Increased uptake of behaviours \\
    d & To persuade & Altered behavioural intentions\\
      & & Increased uptake of behaviours \\
    e & To provide comfort & Reduced psychological distress \\
      & & Reduced anxiety\\
    \midrule
    \# & Effect & \\
    \midrule
    \multicolumn{3}{l}{Communicate information more effectively}\\
    1 & \multicolumn{2}{l}{More engaging} \\
    2 & \multicolumn{2}{l}{Better recall} \\
    3 & \multicolumn{2}{l}{Develop fewer counter--arguments}\\
    \multicolumn{3}{l}{Change attitudes, judgements and behaviours}\\
    4 & \multicolumn{2}{l}{Increase attitudes} \\
    5 & \multicolumn{2}{l}{Reduce prejudice} \\
    6 & \multicolumn{2}{l}{Promote positive behaviour} \\
    7 & \multicolumn{2}{l}{Reduce negative behaviour} \\
    8 & \multicolumn{2}{l}{Improve work performance} \\
    9 & \multicolumn{2}{l}{Ignore base rate information} \\
      & \multicolumn{2}{l}{and increase narrative information}\\
  \bottomrule
\end{tabular}
\end{table}

\subsection{Challenges to software engineering}

Like all disciplines, software engineering faces fundamental challenges. We very briefly consider some of those challenges here, emboldened in the following discussion. In Section \ref{section:contribution-of-storytelling}, we suggest that storytelling can help us to make progress on at least some of these challenges.

Software engineering practice and research is \textbf{multidisciplinary}, integrating many different disciplines from social science through to discrete, and now quantum, mathematics. There is the challenge of respecting the contrasting ways of knowing and types of knowledge from these disciplines. Evidence Based Software Engineering (EBSE) aims ``\dots to improve decision making related to software development and maintenance by integrating current best evidence from research with practical experience and human values'' (\cite{dyba2005evidence}, p. 59). To these we may add system constraints. These different elements -- research, practical experience, human values and constraints -- imply different kinds of knowledge, and there is the challenge of \textbf{integrating different kinds of knowledge}. The \textbf{context} of software practice undermines our ability to generalise findings from research. Context also introduces postmodernist perspectives into software engineering research, e.g., that there is no objective truth at the levels of reality relevant to software engineering. Practitioners and researchers value different kinds of \textbf{evidence}. Different kinds of evidence align with different ways of knowing and support different kinds of knowledge. This introduces the challenge of persuading software practice to improve on the basis of research. There is increasing \textbf{methodological diversity} bringing a challenge to evaluating the quality of research. Finally, there is the challenge of \textbf{retaining, even honouring, meaning} in software engineering in the face of abstraction. In two recent tweets, Michael ``GeePaw'' Hill writes the following: ``The software trade has a great many problems, but among the most debilitating and dangerous is the steadfast refusal to adequately incorporate the humanity of the makers into its culture, organization, and reasoning.'' \cite{hill2021a}, and, ``By a process of relentless abstraction and ruthless compartmentalization, we seek time and again to suppress, distract from, and minimize the most central fact of software development: humans make software.'' \cite{hill2021b}. 

\subsection{Story and storytelling in research}
\label{subsection:prior-research-on-story}
Storytelling is recognised as a legitimate focus of study in other scientific disciplines, e.g., energy and climate change research \cite{moezzi2017using}, law \cite{anderson2005analysis, twining1994rethinking}, organisational research \cite{weick1995sensemaking}, and behavioural medicine \cite{shaffer2018usefulness}. And whilst storytelling is recognised in a variety of disciplines related to software engineering -- e.g., information systems \cite{Schwabe2019specialissue}, human--computer interaction \cite{blythe2017research}, computer supported cooperative work \cite{orr1986narratives}, information visualisation \cite{gershon2001storytelling}, multimedia systems \cite{lugmayr2017serious} and computer science education \cite{kelleher2007using, naul2020story} -- software engineering practice and research has tended to direct attention only at one particular type of \textit{story}, i.e., the user story. 

In behavioural medicine, Shaffer et al. \cite{shaffer2018usefulness} cite prior work to assert several benefits with narrative--as--story. But Shaffer et al. \cite{shaffer2018usefulness} also write that, ``\dots interventions with narratives appear to escape the scrutiny that interventions with statistical evidence receive from people who disagree with the message. This has important implications for health behavior change interventions [and, for the current paper, there are important implications for interventions in software engineering practice], where the goal is often to change attitudes towards an unhealthy or harmful health behavior [or, for the current paper, the goal of improving a suboptimal or counterproductive software practice].'' (\cite{shaffer2018usefulness}, p. 431). We return to Shaffer et al.'s work \cite{shaffer2018usefulness, shaffer2013all} later in this paper.

There is some attention in software engineering research directed at narrative analysis and synthesis (e.g., \cite{cruzes2015case}) however that work uses the term \textit{narrative} in the sense of the qualitative inquiry into and with texts rather than as \textit{story}. We find several papers that explicitly recognise story and storytelling in software engineering research: Ahonen and Sihvonen's \cite{ahonen2005things} story of software process improvement; Lenberg et al.'s unpublished manuscript \cite{lenberg2017behavioral} on guidelines for qualitative studies of behavioural software engineering; and several papers (e.g., \cite{lutters2007revealing,sim2011getting}) that study war stories using storytelling as a method of data collection. We consider these publications in the next section of the paper.

%% file: sections/3_contribution_of_storytelling.tex
\section{The contribution of storytelling to human--centric SE research}
\label{section:contribution-of-storytelling}
% \textcolor{magenta}{Rename as Contribution of S to SE?}

We present several arguments below for why and how storytelling can contribute to software engineering research and practice.
% address at least some of the challenges identified in Section \ref{}.

% \subsection*{Notes}

% \begin{enumerate}
%     \item 
%     \item Somewhere in here combine story with argument and evidence
%     \item Sketch where/how story might fit in SE practice and research
%     \begin{enumerate}
%         \item As method/s of data collection, i.e., collect stories cf. war stories and story stem
%         \item As analyses \& ways of thinking cf. Sim \& colleague
%         \item As complementing argument cf. Rainer, and Law
%         \item As output, cf. Ahonen paper, and quote Lutter and Seaman, and Sim again, and cite others
%         \item As way of knowing \& knowledge, cf. Heron
%         \item As change mechanism, cf. Shaffer et al.
%         \item As advocacy
%     \end{enumerate}
%     \item As threats to validity
%     \begin{enumerate}
%         \item It is dangerous. But so is data cf. AI/ML and fairness etc.
%         \item Good: raises awareness of potential threats
%         \item Protocols (cf. law) and integration of story with argument and evidence (law, Bex et al., etc)
%         \item As acceptable confabulation, cf. Parnas
%     \end{enumerate}
% \end{enumerate}

\subsection{Storytelling as collected data}
Lutters and Seaman \cite{lutters2007revealing} developed a simple protocol for collecting war stories from SE practitioners. That protocol guided practitioners on how to tell their stories. Lutters and Seaman \cite{lutters2007revealing} already demonstrate that software engineering research collects some kinds of story from software practitioners. The stories collected have been stories of exception, e.g., of something that has gone wrong. What constitutes an exception will vary from practitioner to practitioner, e.g., by definition an expert experiences different kinds of exception to a novice. Shaffer et al. \cite{shaffer2018usefulness} show that non--exceptional stories, e.g., of process, can also be valuable. Sim and Alspaugh \cite{sim2011getting} show that collecting war stories is not necessarily just about acquiring a text for analysis. So although stories have been collected in software engineering research, there is much greater opportunity for collecting stories than has been undertaken to date.

\subsection{Storytelling as analysis}
\label{subsection:storytelling-as-analysis}
Sim and Alspaugh \cite{sim2011getting} show that software engineering research has, to date, constrained the way it analyses stories and therefore limited the opportunities for learning from stories. They demonstrate richer ways of analysing stories, drawing on the humanities. They provide example approaches that, due to space, we are unable to review here.
% In law, Anderson \cite{anderson2005analysis} shows how storytelling can help identify gaps in evidence and generate hypotheses.

\subsection{Storytelling and ways of knowing}
There are many different classifications of knowledge and therefore of ways of knowing. We use Heron's \cite{heron1996co} four ways of knowing, briefly summarised here in Table \ref{table:heron_four_ways_of_knowing}. For a detailed exploration of these four ways of knowing, see Heron's book, \textit{Co--operative inquiry: research into the human condition} \cite{heron1996co}. A research field such as software engineering research that is, or seeks to be, multidisciplinary (e.g., \cite{sim2001beg}) must be open to different kinds of knowledge, and therefore different ways of knowing, and not only open to different propositional knowledge.
% Heron constructs four models using these four ways of knowing: a pyramid of knowing, a circuit of knowing, a dialectical process of knowing, and an inquiry cycle. Due to space constraints we do not discuss those here.

\subsubsection{Grounding ways of knowing} 
In Table \ref{table:heron_four_ways_of_knowing}, the four ways of knowing are ordered, with experiential knowing positioned `lowest' and practical knowing positioned 'highest'. This positioning is because the higher--positioned ways of knowing are grounded in the lower--positioned ways of knowing. In software engineering research, for example, we `ground' our propositional knowledge in empirical evidence, and we seek to ground recommendations to practitioners (cf. practical knowledge) in propositional knowledge.
% Heron argues that we must go both down and up the `pyramid' of knowledge, i.e., both grounding upper levels of knowledge in the lower levels, but also fulfilling the lower levels through the upper levels.
% For example, ``Nothing is as practical as a good theory'' (attr. Kurt Lewin)

\subsubsection{Intuition and presentational knowing} Heron's description of presentational knowing as ``\dots an intuitive grasp of the significance of patterns'' might misleadingly suggest that presentational knowing is an unconscious or semi--conscious activity. Storytelling is a form of conscious, intentional presentational knowledge. This is of course most obvious with published novels.

\subsubsection{Assumptions about the status of ways of knowing} Heron is challenging at least two assumptions, i.e.,  1) that propositional knowledge should be the pre--eminent way of knowing, and 2) that propositional knowledge is self--sufficient. Sims et al. \cite{sim2011getting,sim2008marginal} discuss methodical and amethodical knowledge, also challenging the pre--eminence of any particular kind of knowledge.

\subsubsection{The status of experiential and presentational ways of knowing in SE research} It is not clear whether these two types of knowing are distinctly recognised in software engineering research. When we conduct interviews, surveys, focus groups etc. we appear to be accessing presentational ways of knowing. It is frequently hard to directly access the experiential knowledge of others.
% One way to do so would be participant--observation. Another way can be through carefully crafted stories and storytelling.

\begin{table}
  \caption{Heron's \cite{heron1996co} four ways of knowing}
  \label{table:heron_four_ways_of_knowing}
  \begin{tabular}{l p{12cm}}
    \toprule
    Way & Brief description\\
    \midrule
    Practical & `\dots knowing how to exercise a skill\dots'' (p. 52)\\    
    Propositional & ``\dots intellectual statements, both verbal and numeric, conceptually organized in ways that do not infringe the rules of logic and evidence. Propositional knowledge is regarded both as pre--eminent and self--sufficient.'' (p. 33)\\
    Presentational & ``\dots an intuitive grasp of the significance of patterns as expressed in graphic, plastic, moving, musical and verbal art--forms\dots'' (p. 52), e.g., story as presentational knowing.\\
    Experiential & ``\dots imaging [not imagining] and feeling the presence of some energy, entity, person, place, process or thing.'' (p. 52)\\
  \bottomrule
\end{tabular}
\end{table}

% \begin{table}
%   \caption{Heron's \cite{heron1996co} four ways of knowing}
%   \label{table:heron_four_ways_of_knowing}
%   \begin{tabular}{p{8cm}}
%     \toprule
%     Way \& Brief description\\
%     \midrule
%     \textbf{Practical:} `\dots knowing how to exercise a skill\dots'' (\cite{heron1996co}, p. 52)\\    
%     \textbf{Propositional:} ``\dots intellectual statements, both verbal and numeric, conceptually organized in ways that do not infringe the rules of logic and evidence. Propositional knowledge is regarded both as pre--eminent and self--sufficient.'' (\cite{heron1996co}, p. 33)\\
%     \textbf{Presentational:}  ``\dots an intuitive grasp of the significance of patterns as expressed in graphic, plastic, moving, musical and verbal art--forms\dots'' (\cite{heron1996co}, p. 52)\\
%     \textbf{Experiential:} ``\dots imaging [not imagining] and feeling the presence of some energy, entity, person, place, process or thing.'' (\cite{heron1996co}, p. 52)\\
%   \bottomrule
% \end{tabular}
% \end{table}

% \textcolor{magenta}{Also, methodical and amethodical}.

\subsection{Integrating storytelling with evidence and argument for evaluation and assessment}
\label{subsection:story-evidence-argument}
% \textcolor{magenta}{Anderson quote and then my work, and Bex et al.}

Anderson et al. \cite{anderson2005analysis} write: ``\dots in factual enquiries\dots for a story to be accepted as true it needs to be warranted by (anchored in) evidence. A well--informed story needs to be coherent, but to be true, it must be both plausible and backed by particular evidence.'' (\cite{anderson2005analysis}, p. 283)

% Anderson et al. \cite{anderson2005analysis} write: ``In ordinary life and in making important decisions we need stories in order to `make sense' of the world and of particular past events; in factual enquiries, including adjudication, for a story to be accepted as true it needs to be warranted by (anchored in) evidence. A well--informed story needs to be coherent, but to be true, it must be both plausible and backed by particular evidence.'' (\cite{anderson2005analysis}, p. 283)
% Plausibility is tested by background generalizations; the truth of specific factual conclusions is tested by reasons from particular evidence.

Rainer \cite{rainer2017using} developed a preliminary methodology for identifying, extracting and analysing arguments, evidence and explanations from texts, and for presenting those in a structured, integrated way. One type of explanation was the story. He demonstrated the application of the methodology to several examples taken from a blog post by Joel Spolsky \cite{Spolsky2006}, these examples being `micro--war stories'. The methodology seeks to address Anderson's position, e.g., to evaluate the story using evidence. The methodology is open to the limitation that Sim and Alspaugh \cite{sim2011getting} observed with Lutters and Seaman's \cite{lutters2007revealing} approach, i.e., it extracts facts and information from a text. There is therefore the opportunity to extend the methodology. For example, there are opportunities to use story to help evaluate and assess software engineering. Anderson \cite{anderson2005analysis} shows how storytelling can help identify gaps in evidence and can help generate hypotheses.
% (Another significant challenge with the methodology is that, like many methods of qualitative analyses, the methodology is a resource--intensive, manual process.)

% The preliminary methodology was initially based on a method presented by Fisher \cite{fisher2004logic} and then extended to accommodate the wider set of elements present in the framework. One significant challenge with the methodology is that, like many methods of qualitative analyses, the methodology is a resource--intensive, manual process. Although it is a preliminary methodology, and has yet to be further extended or applied, it demonstrates a formal approach for software engineering research to integrate propositional knowledge with story. One direction in which the methodology could be extended is to accommodate the insights of Shaffer et al. \cite{shaffer2018usefulness} into the methodology.

\subsection{Storytelling as output}

There appear to be very few examples where SE researchers publish their research output \textit{as a story}. Ahonen and Sihvonen's \cite{ahonen2005things} paper
% , entitled \textit{How things should not be done: a real--world horror story of software engineering process improvement},
is the only complete example we can find. Ahonen and Sihvonen \cite{ahonen2005things} present a real--world story over a 2.5yr period, told from the point of view of an individual software engineer, of a Software Process Improvement (SPI) effort. They write, ``The story\dots is based mainly on the personal experiences of a single software engineer. Those experiences have been documented by the engineer\dots, but those experiences have [also] been checked by interviewing several other people\dots'' and, ``The reason why a story like this should be interesting for others is that the mistakes made in SPI efforts during the documented time are quite universal.''

% \textcolor{magenta}{Ahonen and Sihvonen's \cite{ahonen2005things} example would be appear to be classified in \dots}

There are many examples of where researchers present fragments of story, e.g., Sim and Alspaugh \cite{sim2011getting, lutters2007revealing}. One explanation for the limited recounting of stories is the amount of publication `real estate' they require.

\subsection{Storytelling as intervention}
\label{subsection:story-intervention}

To explain how storytelling can act as an intervention we draw on Shaffer et al.'s \cite{shaffer2018usefulness} explanatory model. We choose this model because it has been developed by and for scientists.
% It is unfortunate, for the purposes of our discussion, that Shaffer et al. entitled their model the Narrative \textit{Immersion} Model. We (and, we believe, Shaffer et al.) are more interested in the \textit{impact} of a narrative. Immersion appears to be a necessary condition for impact. (The decontextualised, abstracted knowledge often produced by science seems, by definition, to be the opposite of immersion.)

% \subsubsection{Overview}
Shaffer et al. \cite{shaffer2018usefulness} developed the Narrative Immersion Model (NIM) to better understand how storied narrative works so that storied narrative might be used to help behaviour change, e.g., in patients. The NIM may therefore be understood as a model to support intervention. Whilst an intervention might occur \textit{after} the research has taken place
% (e.g., the NIM provides a mechanism for communicating the output of research) 
it is often the case that we design for intervention, e.g., design our studies with the intention of using the results to improve software practice. One implication is that researchers consider storytelling as part of the design of the study.
% \textcolor{magenta}{What advice is their on reporting results, Jedlishka).}

% \subsubsection{Types of story}
Drawing on their prior work \cite{shaffer2013all}, Shaffer et al. \cite{shaffer2018usefulness} define three types of story (see Table \ref{table:how-story-works}). These types are not intended to be exclusive: an actual story might fit more than one type.  \textit{Outcome} stories describe how a situation ends. \textit{Process} stories describe how a decision was made. \textit{Experience} stories describe what a real--world situation was really like. Each type of story has a different effect on behaviour.  Depending on the kind of effect the researchers seek  (cf. Table \ref{table:how-story-works}), the researchers might design for different kinds of story.

% Two example applications of storied narrative, and the NIM, into software engineering research would therefore be around interventions and impact, e.g.,

\begin{table}
  \caption{Shaffer et al.'s types of story \cite{shaffer2018usefulness, shaffer2013all}}
  \label{table:how-story-works}
  \begin{tabular}{l p{8cm}}
    \toprule
    Types & Effect\\
    \midrule
    Outcome & Ability to persuade\\
      & Change attitudes \\
      & Alter intentions \\
      & Alter behaviours \\
     Process & Identify relevant decision attributes\\
      & Model decision processes\\
      & Change how people search for information \\
    Experience & Reduce affective forecasting errors\\
      & Facilitate more accurate perspective--taking\\
      & Improve resilience\\
  \bottomrule
\end{tabular}
\end{table}

Comparing Table \ref{table:how-story-works} with Table \ref{table:purpose_outcome_effects_of_story}, notice that the three types of story identified by Shaffer et al. \cite{shaffer2018usefulness} in Table \ref{table:how-story-works} do not seem to have a direct impact on improving knowledge; or in other words, these stories do not seem to have the function to inform. This may be because Shaffer et al. have not had the opportunity to investigate this outcome.  We make this observation because software engineering research often focuses on knowledge and, more specifically, on propositional knowledge. Storytelling does not seem to naturally fit that type of knowledge, but instead storytelling complements it.

% \subsubsection{Magnitude of effect}
As well as talking about types of story and their effects, Shaffer et al. \cite{shaffer2018usefulness} also talk about the magnitude of effect of a story. Shaffer et al. \cite{shaffer2018usefulness} propose a continuum through which a reader may `travel' from Interest through Involvement to Immersion. For Shaffer et al., the deeper into the narrative a reader `travels' the more powerful the narrative will be to influence behaviour. Characteristics of the narrative promote deeper involvement. These characteristics
% Shaffer et al.'s \cite{shaffer2018usefulness} work, and their
appear to align the NIM with elements of storytelling (cf. Section \ref{section:foundations}) and with advice on creative writing.
% In simple terms, Coyne \cite{coyne2015story} (see also, e.g., \cite{storr2020science}), appear to describe the characteristics that Shaffer et al.'s \cite{shaffer2018usefulness} model needs for immersion.

% In their paper, Shaffer et al. \cite{shaffer2018usefulness} do not explicitly list the characteristics. We therefore briefly summarise some of the characteristics here:

% \begin{itemize}
%     \item Narratives with a greater degree of realism tend to generate more interest, though realism does not mean real.
%     \item A greater connection between the (perceived) narrator and the reader increases the effectiveness of the story.
%     \item A greater connection between the characters in the story and the reader increases the effectiveness of the story. Connection can be increased through (amongst other characteristics) perceived similarity, through liking, and through parasocial interaction. Cultural similarities can also increase persuasion, e.g., through increased trust in the characters.
%     \item Similarity leads to greater 'transportation' (immersion) which requires fewer cognitive resources of the reader to process the story.
%     \item A more coherent story tends to be more persuasive, e.g., that depict causally related events, are temporally ordered and are demarcated by a beginning, middle and end should lead to greater immersion.
%     \item First--person narratives should lead to greater immersion.
%     \item Humour and surprise should increase immersion. For example, a repeating plot structure but with a unexpected `twist' increases surprise.
% \end{itemize}

\subsection{Storytelling as advocacy}
\label{storytelling-as-advocacy}
Stories and storytelling provide a way to advocate for, and/or to raise awareness of, important issues such as human values.  Consider a software engineering team developing a software system for managing information about children in state care homes. What insights might Sissay's \cite{sissay2019my} memoir, of his life in managed care, offer to that software engineering team to help ensure they develop a human--centric software system, and not simply develop an algorithmic bureaucracy? Or, as another example, what insights might Kafka's \textit{The Trial} and \textit{The Castle} provide on AI--based decision--making (e.g., \cite{beck2012weber})? Other examples include: Ford's recently published memoir, \emph{Think Black}~\cite{ford2021thinkblack}, about his father, John Stanley Ford,  hired by Thomas J. Watson to become IBM's first black software engineer;  Wiener's memoir, \emph{Uncanny Valley}~\cite{Wiener2020uncanny}, describing her experiences working in technology companies in Silicon Valley; and Kim's two fictional accounts, one co--written with colleagues, about DevOps \cite{kim2018phoenix} and software development \cite{kim2019unicorn}. 
% entitled \textit{Anaya's journey: a vision for a future software academy},
% advocating for a fictional academy \cite{rainer2020anaya}. The intention with the story was to advocate for a particular kind of higher education of software engineering that was inclusive.

\subsection{The risks of storytelling}
Anderson et al. \cite{anderson2005analysis} write, ``\dots story telling is vulnerable to abuse. It may be true that stories and storytelling are psychologically necessary to decision--making in legal contexts, but they are dangerous in that they often can be used to violate logical standards, appeal to emotion rather than reason, and subvert legal principles and conventions.'' (\cite{anderson2005analysis}, p. 280). Anderson et al. \cite{anderson2005analysis} caution about the careful use of story but do not argue \emph{against} the use of story. They list a number of examples of dangers with story. They also develop a `rough working protocol' for assessing the plausability, coherence and evidentiary support for a story.

All research methods have strengths and weaknesses and the researcher seeks to increase their awareness of the threats to validity that arise with each method. One contribution of storytelling is to help researchers remain aware of threats. Such threats don't just occur in the telling of stories. For example, Sim and Alspaugh \cite{sim2011getting} observe that storytelling is performance: interviewees select and filter what and how they share information. Such selection and filtering doesn't just occur with the telling of a story, however. Any interview can be understood as a performance, with the threats to validity that might arise.

A highly--influential paper in software engineering research, Parnas' \cite{parnas1986rational} \textit{A rational design process: how and why to fake it}, explicitly recognises the value of `faking' something: ``We will never find a process that allows us to design software in a perfectly rational way. The good news is that we can \textit{fake it}. We can \emph{present} our system to others as \emph{if we} had been rational designers and it pays to \textit{pretend} [to] do so during development and maintenance. (\cite{parnas1986rational}, p. 251; emphasis added). Parnas explains the scope of the pretence and argues for its value in certain circumstances. In a fundamental way, abstraction is also pretence, for it presents a decontextualised and therefore simplified version of reality. We accept that pretence because there is value in doing so. We suggest that storytelling shouldn't be summarily dismissed on the basis of pretence.

% As noted earlier, Rainer published a short story describing a fictional academy \cite{rainer2020anaya}. One interesting reaction to an earlier, unpublished version of the story was the following comment from a reader: ``I really thought for a minute this [the Academy] was some existing entity in some unnamed university in Ireland, hence me googling it, and finding `it' which is not it.'' . 
% Storytelling can be dangerous because it can mislead and can do so in subtle ways. The same concerns arise for artificial intelligence. As with artificial intelligence, the argument should not be that storytelling shouldn't be used, but about the responsible use of storytelling.

%% file: sections/4_discussion.tex
\section{Discussion}
\label{section:discussion}

% \subsection*{Notes}
% \begin{enumerate}
%     \item Many things we haven't talked about here
%     \begin{enumerate}
%         \item User story, persona etc
%         \item Research in distant disciplines -- quote some
%         \item Research in nearby disciplines -- quote some
%         \item
%     \end{enumerate}
%     \item Many things still to be worked out
%     \begin{enumerate}
%         \item Resistance to change
%     \end{enumerate}
% \end{enumerate}

\subsection{Where next?}

There are many aspects of storytelling in human--centric software engineering that we have not discussed in this paper, and further research is needed. Areas requiring further attention include:

\begin{enumerate}
    \item A systematic review of storytelling that builds on the brief review presented in Section \ref{subsection:prior-research-on-story}. Such a review should consider both the humanities and the sciences, as well as creative writing and `story analysts', e.g., \cite{coyne2015story, storr2020science}.
    \item The development of conceptual frameworks for storytelling that build on the foundations introduced in Section \ref{section:foundations}. Included within this framework would be the development of appropriate terminology, e.g., \textit{argued storytelling} or \textit{evidence--based storytelling} are phrases that could communicate the intent of the kind of storytelling we consider here.
    % and, importantly, that seeks to integrate different perspectives on story.
    \item The development of research methodology and methods that draw on work from other disciplines.
    % Such methodology would extend across the full research life--cycle from study design through to dissemination and impact. 
    Several examples have been discussed in this paper, e.g., \cite{shaffer2018usefulness,lutters2007revealing,sim2011getting,anderson2005analysis,rainer2017using,tang2016making}.
    
    % e.g., Shaffer et al.'s \cite{shaffer2018usefulness} Narrative Immersion Model, Lutter and Seaman's \cite{lutters2007revealing} War Stories, Sim and Alspaugh's \cite{sim2011getting} variant on the War Story, Anderson's \cite{anderson2005analysis} preliminary protocol for handling stories in law, Rainer's AXE methodology \cite{rainer2017using}, and Tang et al.'s \cite{tang2016making} Story Stem technique could all provide foundations for methods and methodology in this area.
    \item Guidance on the use of methodology and on assessing the quality of storytelling research, cf. \cite{ralph2021empirical}.
    % to methodology and methods, e.g., an addition to the emerging Empirical Standards in Software Engineering \cite{ralph2021empirical}.
    \item Developing ways in which stories in SE can be evaluated and assessed, but also how storytelling might be used to evaluate and assess software engineering. For example, Section \ref{subsection:storytelling-as-analysis} briefly discussed the use of storytelling to generate hypotheses, Section \ref{subsection:story-evidence-argument} discussed the integration of storytelling with argument and evidence, and Section \ref{subsection:story-intervention} discussed the use of storytelling for intervention.
    \item Investigating the use of storytelling as an approach to knowledge exchange with industry, and as a method of appropriate intervention in practice, cf. \cite{shaffer2018usefulness}.
    \item Investigating the benefits of a more sophisticated understanding of storytelling, beyond the user story, for software engineering practice.
    \item The use of both reading groups and writing groups for developing skills in appreciating stories in research and in practice. 
    % Consider a software engineering team developing a software system for managing information about children in state care homes. What insights might Sissay's \cite{sissay2019my} memoir, of his life in managed care, offer to that software engineering team to help ensure they develop a human--centric software system, and not simply develop an algorithmic bureaucracy? Or, as another example, what insights might Kafka's \textit{The Trial} and \textit{The Castle} provide on AI--based decision--making (e.g., \cite{beck2012weber})? 
    Section \ref{storytelling-as-advocacy} presented  examples that reading groups might use.
    \item The development of platforms, resources and (social) networks to disseminate work, including supplementary materials. Examples to draw on include StoryCorps \cite{StoryCorps2021}, The Story Collider \cite{StoryCollider2021} and experience repositories \cite{schneider2003effective}.

    % \item Focus here on research, but at least some applies to practice
    % \begin{itemize}
    %     \item Steps for research
    %     \item Steps for practice
    % \end{itemize}
    % \item Develop frameworks or models cf. AXE
    % \item Developing new skills
    % \item Reading groups
    % \item Writing groups cf. Faber Academy
    % \item Storytelling platforms \& resources
    % \begin{itemize}
    %     \item Network
    %     \item StoryCorps
    %     \item Agile Corps
    %     \item Story Collider
    % \end{itemize}
    % \item Dissemination of \textit{argued stories} and \textit{evidence--based storytelling}
    % \begin{itemize}
    %     \item Conference workshop/s
    %     \item Sections in journals 
    % \end{itemize}
    % \item Change to metrics of impact
    % \item Standards for acceptable empirical research in SE
    % \item Multidisciplinary collaboration
    % \item Supplementary materials (papers as code, papers as story)
    % \begin{itemize}
    %     \item cf. PROMISE
    %     \item papers with code
    % \end{itemize}
    % \item Methodology
    % \begin{enumerate}
    %     \item War story interview
    %     \item Story Stem Technique
    %     \item Protocol/s for handling stories
    %     \begin{enumerate}
    %         \item Argument schemes
    %         \item Legal protocols
    %     \end{enumerate}
    %     \item Integrating story with other kinds of knowing
    % \end{enumerate}
\end{enumerate}

\subsection{Conclusion}
In this paper, we position storytelling as a particular kind of narrative. We show that whilst narrative synthesis is recognised in software engineering, storytelling has not been properly considered. We explore types of story and the outcomes and effects of these different story types. We argue that storytelling has a valuable, even necessary, contribution to many areas of human--centric software engineering. We recognise that stories can be dangerous (e.g., because they can mislead) but also that these dangers might be assessed and mitigated. We suggest that storytelling may act as a potential counter--balance to abstraction, and a means to retain and honour meaning in software engineering.